\documentclass[aps,amsmath,amssymb, notitlepage, twocolumn,nofootinbib,superscriptaddress,longbibliography]{revtex4-1}
\usepackage{braket}
\usepackage{amsmath,amsfonts, amssymb, amsthm, dsfont,mathtools,breqn}
\usepackage{yfonts}
\usepackage{bm}
\usepackage{mathrsfs}
\usepackage{graphicx}
\usepackage{verbatim}
\usepackage{color}

\newcommand{\floor}[1]{\lfloor #1 \rfloor}
\renewcommand{\refeq}[1]{Eq.~\eqref{#1}}

\begin{document}
\title{Short Proofs of Linear Growth of Quantum Circuit Complexity}
\author{Zhi Li}\affiliation{\PI}
\newcommand*{\PI}{Perimeter Institute for Theoretical Physics, Waterloo, Ontario N2L 2Y5, Canada}

\begin{abstract}
The complexity of a quantum gate, defined as the minimal number of elementary gates to build it, is an important concept in quantum information and computation. It is shown recently that the complexity of quantum gates built from random quantum circuits almost surely grows linearly with the number of building blocks. In this article, we provide two short proofs of this fact. We also discuss a discrete version of quantum circuit complexity growth.

\end{abstract}
\maketitle

\section{Introduction}
\newcommand{\bUB}[1]{\overline{\mathcal{U}^B(#1)}}
\newcommand{\UB}[1]{\mathcal{U}^B(#1)}

The complexity of a quantum gate, defined as the minimal number of elementary gates to build it, is an important concept in quantum information and computation \cite{doi:10.1137/S0097539796300921} and finds interesting applications in high energy physics \cite{PhysRevD.90.126007}. It has been conjectured that typical local random circuits produce quantum gates with linearly growing  (in time) complexity for an exponentially (in system size) long time \cite{PhysRevD.97.086015}.

Recently, Ref.~\cite{naturepaper} proved a version of the above conjecture. Since all $n$-qubit gates can be synthesized by 2-qubit gates drawn from $\mathbb{SU}(2^2)$, we define the complexity $\mathcal{C}(g)$ of a $n$-qubit gate $g\in \mathbb{SU}(2^n)$ as the number of 2-qubit gates in a minimal synthesis. Now assuming we build $n$-qubit gates by concatenating some building blocks $B$, each containing $|B|$ arbitrary 2-qubit gates arranged in a fixed structure. Then almost surely (with probability 1) and before a time scale exponential in $n$, $\mathcal{C}(g)$ grows linearly with the number of building blocks. 

In this article, we provide two short proofs of this fact.

\section{First Proof}

We denote the space of $n$-qubit gates after $k$ blocks as $\mathcal{U}^B(k)$. It is the image of a construction map 
\begin{equation}\label{eq-Fk}
    F_k: \mathbb{SU}(4)^{k|B|}\to\mathbb{SU}(2^n).
\end{equation}
$\mathcal{U}^B(k)$, as a semialgebraic set, can be decomposed as disjoint union of smooth manifolds, and has a well-behaved dimension theory \cite{bochnak2013real,naturepaper}. Denote $d^B(k)$ to be its dimension.

Define $K$ to be the minimal integer such that $\bUB{K}=\mathbb{SU}(2^n)$. Here the closure is taken in the Zariski topology (regarding $\mathbb{SU}(2^n)$ as an affine variety in $\mathbb{R}^{2^{2n+1}}$). The (algebraic or geometric) dimension of $\bUB{k}$ is the same as the dimension of $\UB{k}$ \cite{bochnak2013real}. By counting degrees of freedom,
\begin{equation}\label{eq-upperbound1}
    d^B(k)\leq 15k|B|.
\end{equation} Hence $K$ is at least exponential in $n$.

We claim that $\UB{k+1}\nsubseteq\bUB{k}$ for $\forall k<K$. If not, assuming $\UB{k_0+1}\subseteq\bUB{k_0}$ for some $k_0<K$, % $\bUB{k}\subseteq\bUB{k+1}\subseteq\bUB{k}$ hence $\bUB{k+1}=\bUB{k}$.
we have
\begin{equation}
   \bUB{k_0} \subseteq\bUB{k_0+1}\subseteq\bUB{k_0} ~\Rightarrow~  \bUB{k_0}=\bUB{k_0+1},
\end{equation}
and\footnote{The second $\subseteq$ is valid for all subsets $X, Y$ in a topological group: 
    $X\overline{Y}=\bigcup_{x\in X}x\overline{Y}=\bigcup_{x\in X}\overline{xY}\subseteq \bigcup_{x\in X} \overline{XY}=\overline{XY}$.}
\begin{equation}
\begin{split}
    \UB{k_0+2}&=\UB{1}\UB{k_0+1}\subseteq \UB{1}\bUB{k_0} \\
    &\subseteq \overline{\UB{1}\UB{k_0}}=\bUB{k_0+1}.
\end{split}
\end{equation}
Iteratively, we get
\begin{equation}
    \bUB{k_0}=\bUB{k_0+1}=\bUB{k_0+2}=\cdots=\bUB{K}.
\end{equation}
This contradicts the definition of $K$.

This means that, for $\forall k<K$, there exists a $g\in\UB{k+1}$ such that $g\notin \bUB{k}$. By definition of the Zariski topology, there exist a polynomial $p$ on $\mathbb{SU}(2^n)$ such that $p=0$ on $\UB{k}$ but $p(g)\neq 0$. By definition of $\UB{k+1}$, 
\begin{equation}
 \exists x\in\mathbb{SU}(4)^{(k+1)|B|},~p(F_{k+1}(x))\neq 0.
\end{equation}
where $F_{k+1}$ is the construction map in \refeq{eq-Fk}. 

$p\circ F_{k+1}$ being a nonzero polynomial on $\mathbb{SU}(4)^{(k+1)|B|}$, its zero point set must have zero measure. Hence for almost all $x$, $p(F_{k+1}(x))\neq 0$ and therefore $F_{k+1}(x)\notin\UB{k}$. In other words, almost all $n$-qubit gates generated with $k+1$ blocks cannot be generated using $k$ blocks.

\textbf{Remark}: This proof can be modified to the state complexity growth by considering the space of states after $k$ blocks.

\section{Second Proof}

This proof uses the observation in Ref.~\cite{naturepaper} that the dimension $d^B(k)$ is a proxy for the complexity. 

We note the following facts:
\begin{itemize}

\item  $d^B(k)$ is non-decreasing in $k$;
\item A $(k_1+k_2)$-block circuit can be decomposed as two circuits with $k_1$ and $k_2$ blocks, hence
\begin{equation}\label{eq-relation}
    d^B(k_1+k_2)\leq d^B(k_1)+d^B(k_2);
\end{equation}
\item $O(2^{2n})$ two-qubits gates  are enough to synthesize all $n$-qubit gates \cite{gate2003}. Therefore, as long as $\mathcal{U}^B(1)$ contains all two-qubit gates,
\begin{equation}\label{eq-enoughgates}
    d^B(2^{2n+c_1})\geq \dim \mathbb{SU}(2^n)=2^{2n}-1.
\end{equation}
Here $c_1$ (and later $c_2$ and $c$) is a constant.
\end{itemize}
From these facts we can deduce a lower bound of $d^B(k)$ for all $k\leq 2^{2n}$:
\begin{equation}
    d^B(k)\geq d^B(2^{r})\geq \frac{d^B(2^{r+1})}{2}\geq\cdots\geq\frac{d^B(2^{2n+c_1})}{2^{2n+c_1-r}}> c_2k.
\end{equation}
Here $r=\floor{\log_2k}$.
% ; $c_2>0$ is a constant.

On the other hand, define $\mathcal{U}(s)$ as the space of $n$-qubit gates built from arbitrary $s$ 2-qubit gates in all possible structures. Denote $d(s)$ to be its dimension. Similar to \refeq{eq-upperbound1}, we know $d(s)\leq 15s$.
Then we have:
\begin{equation}
    d^B(k)> 15\frac{c_2k}{15}\geq d(\floor{ck}).
\end{equation}
This means, at step $k$, those ``short-cut" gates with no larger than $\floor{ck}$ gate complexity is subdimensional in $\mathcal{U}^B(k)$.

Let us decompose $\mathbb{SU}(4)^{k|B|}$ as $R\cup R^c$, where $R$ is the set of regular points where $F$ has maximal rank;  its complement $R^c$ is the set of critical points. Then $F_k^{-1}(\mathcal{U}(\floor{ck}))$ can be decomposed as
\begin{equation}
    (F_k^{-1}(\mathcal{U}(\floor{ck}))\cap R^c)\bigcup(F_k^{-1}(\mathcal{U}(\floor{ck}))\cap R).
\end{equation}
$R^c$ has measure 0 as a subvariety, since the rank can be characterized by minor determinants which are polynomial functions. The second term also has measure 0 since locally they are (unions of) subdimensional manifolds. Therefore,  the preimage of $\mathcal{U}(\floor{ck})$ has measure 0. 

In other words, almost all gates at step $k$ have greater than $\floor{ck}$ gate complexity.

\textbf{Remark}: For fact \refeq{eq-enoughgates} to hold, $|B|=O(n)$ is enough. For example, we can use some SWAP gates to convert a gate acting on qubits $i$ and $j$ to a gate acting on qubits 1 and 2.

\section{Discussions}

Besides synthesizing $n$-qubit gates by 2-qubit gates drawn from $\mathbb{SU}(2^2)$, one can also draw 2-qubit gates from a discrete, universal gate set (for example, Clifford and $T$). In this setting, there is also a version of 
quantum circuit complexity growth.

More precisely, denote $g_k$ as the resulting gate by randomly applying some gates from a fixed universal gate set $\mathcal{S}$ (assuming $\mathcal{S}=\mathcal{S}^{-1}$ without loss of generality) for $k$ times:  $g_k=h_1h_2\cdots h_k$, $h_i\in \mathcal{S}$. Define the complexity $\mathcal{C}(g_k)$ as the minimal number of gates to build $g_k$ from gates in $\mathcal{S}$. Then there exist a constant $c>0$ such that
\begin{equation}\label{eq-Kingman}
    \lim_{k\to\infty}\frac{\mathcal{C}(g_k)}{k}=c \text{~~almost surely}.
\end{equation}
Note that different from the continuous version, the discrete complexity grows forever. 
 
This statement is a known mathematical fact from the study of random walks on groups. A proof sketch is as follows (see Ref.~\cite{zheng2021asymptotic} for a survey on terminologies used here). 
The  complexity  satisfies a subadditive relation similar to \refeq{eq-relation}:
\begin{equation}
    \mathcal{C}(g_{k_1+k_2})\leq  \mathcal{C}(g_{k_1})+ \mathcal{C}(g_{k_2}).
\end{equation}
By Kingman’s subadditive ergodic theorem, $\lim_{k\to\infty}\frac{\mathcal{C}(g_k)}{k}=c$ almost surely, where $c$ is a circuit independent constant ($c\geq 0$ for now). Similarly, $\lim_{k\to\infty}\text{Prob}(g_{2k}=id)^{1/2k}=\rho$  exists. It can be proved that $\rho<1$ implies $c>0$. By Kesten's theorem, $\rho<1$ if and only if $\langle\mathcal{S}\rangle$, the subgroup generated by $\mathcal{S}$, is nonamenable. In our setting, $\langle\mathcal{S}\rangle$ is a dense subgroup of $\mathbb{SU}(2^n)$, so by Tits alternative and Ref.~\cite{breuillard2003dense} (or Ref.~\cite{rosenblatt1974invariant}) it is indeed nonamenable.

It must be emphasized that the above-defined complexities (both the continuous and the discrete version) are exact complexities. The case of the approximate complexity, where gate synthesis can be approximate yet the complexity is still conjectured to grow linearly, is a more difficult question since it requires understanding how gates distribute on $\mathbb{SU}(2^n)$ \cite{PRXQuantum.2.030316}.

    % \item Proving the linear growth for approximate complexity seems to be a much harder question.

% \break
\begin{acknowledgements}
The author thanks Timothy H. Hsieh, Cheng-Ju Lin, Zi-Wen Liu, and Beni Yoshida for helpful discussions. Research at Perimeter Institute is supported in part by the Government of Canada through the Department of Innovation, Science and Economic Development and by the Province of Ontario through the Ministry of Colleges and Universities.
\end{acknowledgements}

\bibliography{ref.bib}
\end{document}